\documentclass[twocolumn,showpacs,preprintnumbers,amsmath,amssymb,
superscriptaddress]{revtex4}
\usepackage{graphicx}% Include figure files
\usepackage{dcolumn}% Align table columns on decimal point
\usepackage{bm}% bold math
\usepackage{bbm}
\usepackage{epsfig}
\usepackage{mathrsfs}
\usepackage{color}
\newcommand{\uqcomment}[1]{}
\def\COMMENTS{
\renewcommand{\uqcomment}[1]{\\\textcolor{red}{\emph{##1}}\\}}
\COMMENTS
\newcommand{\eocomment}[1]{}
\def\COMMENTS{
\renewcommand{\eocomment}[1]{\\\textcolor{blue}{\emph{##1}}\\}}
\COMMENTS
\begin{document}

%%%%%%%%%%%%%%%%%%%%%%%%%%%%%%%%%%%%%%%%%%%%%%%%%%%%%%%%%%%%%%%%%
\title{Quantum effects in the dynamical localization of Bose-Einstein condensates \\ in optical lattices}% Force line breaks with \\

\author{Beata~J.~D\c{a}browska-W\"uster}
\email{bjd124@rsphysse.anu.edu.au}
\affiliation{Australian Research Council Centre of Excellence for Quantum-Atom Optics}
\affiliation{
Nonlinear Physics Centre, Research School of Physical Sciences and Engineering, Australian National University, 
Canberra ACT 0200, Australia}
\author{Sebastian~W\"uster}
\affiliation{Australian Research Council Centre of Excellence for Quantum-Atom Optics}
\affiliation{Department of Physics, Faculty of Science, Australian National University, Canberra ACT 0200, Australia}
\author{Ashton~S.~Bradley}
\affiliation{Australian Research Council Centre of Excellence for Quantum-Atom Optics}
\affiliation{School of Physical Sciences, University of Queensland, Brisbane QLD 4072, Australia}
\author{Matthew~J.~Davis}
\affiliation{Australian Research Council Centre of Excellence for Quantum-Atom Optics}
\affiliation{School of Physical Sciences, University of Queensland, Brisbane QLD 4072, Australia}
\author{Elena~A.~Ostrovskaya}
\affiliation{Australian Research Council Centre of Excellence for Quantum-Atom Optics}
\affiliation{
Nonlinear Physics Centre, Research School of Physical Sciences and Engineering,
Australian National University, Canberra ACT 0200, Australia}

%%%%%%%%%%%%%%%%%%%%%%%%%%%%%%%%%%%%%%%%%%%%%%%%%%%%%%%%%%%%%%%%%


\begin{abstract}

We study quantum effects in the dynamics of a Bose-Einstein condensate loaded onto the edge 
of a Brillouin zone of a one-dimensional periodic potential created by an optical lattice. 
We show that quantum fluctuations trigger the dynamical instability of the Bloch states 
of the condensate and can lead to the generation of arrays of {\em matter-wave gap solitons}. 
Our approach also allows us to study the instability-induced anomalous heating of the condensate 
at the edge of the Brillouin zone and growth of the uncondensed atomic fraction. 
We demonstrate that there are regimes in which the heating effects do not suppress the formation 
of the localised states. 
We show that a phase imprinting technique can ensure the formation of gap soliton
trains after short evolution times and at fixed positions.
\end{abstract}

			     
\maketitle
%
%%%%%%%%%%%%%%%%%%%%%%%%%%%%%%%%%%%%%%%%%%%%%%%%%%%%%%%%%%%%%%%%%
\section{\label{sec:intro}Introduction\protect}

Bose-Einstein condensates (BECs) loaded into optical lattices provide a unique opportunity 
for testing many of the fundamental concepts of solid state physics in a flexible 
and  defect-free model system.
However, one feature sets this system apart from any condensed matter system --- the intrinsic
nonlinearity of the BEC that is fundamentally due to elastic atomic  scattering. 
The physics of the intricate interplay between the periodicity of the lattice and nonlinearity 
of coherent matter wave has recently been subject to numerous theoretical and experimental 
studies, see Refs.~\cite{ol_review,lewenstein_review} for an overview.

One of the most striking manifestations of this interplay observed so far is formation of bright 
atomic solitons in a condensate with {\em repulsive} nonlinear interactions~\cite{Eiermann2004}. 
This unique form of nonlinear localisation occurs within the {\em gaps} of the periodicity-induced 
band-gap spectrum of the matter Bloch waves, and is possible only due to anomalous diffraction 
properties that the matter waves acquire at the edges of the Brillouin zone due to the Bragg 
scattering  on a periodic potential. The experimental observation of a localized excitation 
(gap soliton) required preparation of the condensate with a {\em small number of atoms} 
in the ground state  of the one-dimensional lattice in the middle of the Brillouin zone. 
The BEC was then adiabatically driven to the band-edge in an accelerating lattice, followed by
evolution at the band edge in the lattice moving with a constant velocity. 

Employing a similar procedure, but with condensates containing a large number of atoms, 
Fallani~\emph{et al.}~\cite{Fallani2004} observed another basic nonlinear effect --- the
dynamical (modulational) instability of a Bloch wave evolving at the edge of a Brillouin zone in
a moving lattice.  The observed signatures of this instability were significant loss of atoms
from the condensate and its spatial fragmentation. The loss of atoms is manifested in the 
``anomalous heating'', i.e. enhanced growth of the thermal or uncondensed fraction of atoms. 
On the other hand, theoretical studies of condensates with a large number of atoms at the edge 
of the Brillouin zone beyond the onset of the dynamical instabilities have shown that the
instability-induced dynamics can also result in condensate localization  and formation of a {\em
train} of the localized gap soliton-like structures~\cite{Dabrowska2006}. These studies were
performed in the framework of the mean-field (Gross-Pitaevskii) model, and therefore the loss of
atoms from the condensate and formation of a thermal cloud could not be
captured by this analysis. 

The purpose of this paper is to answer the question: 
{\em Can periodic condensate localization at the band edge still occur in the presence of
anomalous heating?} To this end, we perform an analysis of the condensate 
non-adiabatically loaded into a moving optical lattice using the truncated Wigner method 
for BEC dynamics~\cite{QN,Drummond1987,Steel1998,Sinatra2001,
Castin2002,Lobo2003,Norrie2005,Norrie2006} that incorporates some effects of quantum 
fluctuations, and allows the formation of a non-condensed fraction. 
Using this model we show that, in certain regimes of condensate dynamics, the anomalous 
heating of the condensate {\em does not prevent} its localization at the band edges. 
However, the quantum noise that triggers the instability will lead to different dynamics in
every experiment. We analyse signatures of localisation both, in a single-trajectory realization 
of the quantum field, which is fairly similar to a ``single-shot'' experiment, 
and via ensemble averages.

\section{\label{sec:model}Formalism\protect}
The second-quantized Hamiltonian for a Bose gas of interacting atoms in an external 
trapping potential is given by
\begin{align}
\hat{H}&=\int dx \,\hat{\Psi}^{\dag}\hat{H}_{0}\hat{\Psi}
+\frac{U}{2}\int dx \,\hat{\Psi}^{\dag} \hat{\Psi}^{\dag} \hat{\Psi} \hat{\Psi},
\label{hamiltonian}
\\
\hat{H}_{0}&=-\frac{\hbar^2}{2 m} \nabla^2 + V(x,t),
\nonumber
\end{align}
where $\hat{\Psi}(x)$ is the atomic field operator that annihilates a particle at position $x$, 
$m$ is the atomic mass and $U=4 \pi \hbar^2 a_s/m$ is the interaction strength, where $a_s$ 
is the $s$-wave scattering length. The trapping potential $V(x,t)$ generally depends on both time 
and position. 
Given the Hamiltonian Eq.~(\ref{hamiltonian}), the evolution of the system's density operator 
$\hat{\rho}$ obeys the von-Neumann equation
\begin{align}
\frac{d \hat{\rho}}{d t}=-\frac{i}{\hbar}[\hat{\rho},\hat{H}].
\label{mastereqn}
\end{align}
In general it is not possible to find analytic solutions for Eq.~(\ref{mastereqn}), 
and numerical methods must be used instead.

One manner in which to proceed is to use a phase-space method to represent the density
operator by an expansion on a suitable basis, and then derive equations of motions for the
phase space variables~\cite{QN}.  In principle the positive-P representation~\cite{Drummond1980} 
and its extensions~\cite{deuar2001,deuar2006,corney2004} are exact methods
for quantum dynamics and have had several successes in simulating Bose gas systems.  However,
as these methods are stochastic in nature, in general they suffer from an exponentially
increasing sampling error that can limit the useful simulation time.

In this paper we make use of another phase space technique known as the 
truncated Wigner approximation.  This was used in the context of squeezing 
of solitons in optical fibres by Drummond and Carter~\cite{Drummond1987}, 
and was first applied to Bose gases by Steel~\emph{et al.}~\cite{Steel1998}.
We proceed from the Wigner function $W(\alpha(x),\alpha^*(x))$ for the system 
which is derived from the density operator $\hat{\rho}(x)$ in the single mode case 
via the characteristic function~\cite{QN}:
\begin{align}
W(\alpha,\alpha^*)&=\int e^{\beta^*\alpha-\beta\alpha^*}Tr\{e^{\beta\hat{a}^{\dag} - \beta^*\hat{a}}\hat{\rho}\}\:d^2\beta.
\label{wignerexpansion}
\end{align}
The multimode functional generalization is straightforward, and it is possible to derive 
an evolution equation for $W(\alpha(x),\alpha^*(x))$ in the form
\begin{align}
&\frac{\partial \: W\left(\alpha(x),\alpha^*(x)\right)}{\partial t}=
\label{fokkerplanck}
\\
&  -i \int_{-\infty}^{\infty} \!\! dx \; \frac{1}{\hbar} \left\{\, \frac{\delta}{\delta \alpha(x)}
\left[\hat{H}_{0} \; \alpha(x) - U \left|\alpha(x) \right|^{2} \right] - \; c.c.
\right.
\nonumber
\\
&
\left.
+ \: \frac{U}{4} \:\: \frac{\delta^{3} }{\delta\alpha(x)^{2} \: \delta\alpha^{*}(x)} \; - \; c.c.
\right\} \, W \left( \alpha(x),\alpha^*(x) \right).
\nonumber
\end{align}
If the terms involving third order functional derivatives with respect to $\alpha$
are omitted Eq.~(\ref{fokkerplanck}) takes the form of a Fokker-Planck equation~\cite{QN},
which can then be solved with stochastic methods.
This approach has become known as the \emph{truncated Wigner approximation}~(TWA).
The omission of the higher order terms in Eq.~(\ref{fokkerplanck}) is justified when there 
are more particles than modes in a calculation and the simulated times 
are short~\cite{Castin2002,Norrie2005,Norrie2006,Polkovnikov}.  

With the higher order derivatives neglected, the equation of motion for the stochastic 
wave function $\alpha(x)$ that is equivalent to Eq.~(\ref{fokkerplanck}), has  the same form 
as the usual Gross-Pitaevskii equation~(GPE)~\cite{Drummond1987,Steel1998}:
\begin{align}
i\hbar \frac{\partial \alpha(x)}{\partial t}
=\left(-\frac{\hbar^2}{2 m} \nabla^2 + V + U|\alpha(x)|^2\right) \alpha(x).
\label{sgpe}
\end{align}
The difference from the Gross-Pitaevskii theory arises in the stochastic initial conditions, 
where quantum fluctuations enter the description of the initial physical state. 

In order to obtain quantum correlations of the atomic field $\hat{\Psi}$, 
Eq.~(\ref{sgpe}) has to be solved for a large number of trajectories 
whose initial states $\alpha(x,t=0)$ are assigned according 
to the Wigner distribution of the assumed quantum state of the BEC. 
In this paper we denote quantum ensemble operator averages by angle brackets e.g. 
$\langle \hat{\Psi}^{\dag}(x) \hat{\Psi}(x) \rangle$, and averages
over trajectories using overlines e.g. $\overline{\alpha^*(x)\alpha(x)}.$
In the Wigner representation trajectory averages 
give symmetrically ordered operator averages \cite{QN}, for example
\begin{align}
\overline{ \alpha^*(x,t)\alpha(x,t)} =\frac{1}{2}\langle \hat{\Psi}^\dag(x,t)\hat{\Psi}(x,t)
+\hat{\Psi}(x,t)\hat{\Psi}^\dag(x,t)\rangle.
\label{Wav}
\end{align}
In our numerical implementation of the Wigner method, we extract the condensate density 
and total (condensed and thermal) atomic density by finding the mean-field
\begin{align}
\psi_{0}(x) \equiv \langle \hat{\Psi}(x) \rangle = \langle\alpha(x)\rangle,
\end{align}
\begin{align}
n_{0}(x) \equiv |\psi_{0}(x)|^2,
\end{align}
\begin{align}
n_{\rm tot}(x) & \equiv n_{0}(x) + n_{\rm therm}(x)  
\\
&\equiv \langle \hat{\Psi}^{\dagger}(x) \hat{\Psi}(x) \rangle 
= \overline{ \alpha^{*}(x) \alpha(x) } - \frac{1}{4 h_x},
\nonumber
\end{align}
where $h_x$ is the spacing of the computational grid \cite{vacuum_note}.
The number of condensed or thermal atoms is therefore found as:
\begin{align}
&N_{\rm cond} = \int\! n_{0}(x) \, dx,
\\
&
N_{\rm therm} = \int\! n_{\rm therm}(x) \, dx.
\nonumber
\end{align}

\section{Simulation parameters for soliton formation}  
We consider the quantum dynamics of a strongly anisotropic cigar-shaped BEC cloud loaded 
into a one-dimensional optical lattice. 
Provided that the condensate is tightly confined in the plane transverse to the direction of the lattice,
this problem can be considered to be effectively one-dimensional. 
The dimensionality of the equations of motion can be reduced by applying standard 
procedures~\cite{1d_reduction}, which results in the following 1D model:   
\begin{align}
i \frac{\partial \alpha(x)}{\partial t}
=\left(-\frac{1}{2} \frac{\partial^2}{\partial x^2} + V_{\rm OL}(x,t) 
+ \gamma|\alpha(x)|^2\right) \alpha(x),
\label{1Dsgpe}
\end{align}
where $\gamma$ is the rescaled one dimensional interaction strength $\gamma= 2 a_s/a_{\perp}$ 
and length, time and energy are expressed in units: 
$a_\perp=(\hbar/m \omega_{\perp})^{\frac{1}{2}}$, $\omega_{\perp}^{-1}$ 
and $\hbar\omega_{\perp}$ respectively. 
The confining potential of the time dependent optical lattice is defined 
as $V_{\rm OL}(x,t)=V_0\sin^2(\pi x/d-vt)$, where $V_0$ is the depth, $d=\lambda/2$ 
is the period and $v$ is the velocity of the moving lattice due to non-zero frequency detuning 
between the lattice-forming laser beams.

Throughout this paper we give physical units  of length and time that correspond to a condensate 
of $^{87}\mbox{Rb}$ atoms with atomic mass $m=1.44 \times10^{-25}$kg
confined with the transverse trapping frequency
$\omega_{\perp}=2\pi\times100$Hz, which gives the unit length
of $a_{\perp} \approx 1.08~\mu\mbox{m}$. 
We also set  $\pi/d=1$ which corresponds to an inter-well lattice spacing 
of $\sim 3.4~\mu\mbox{m}$.

We investigate the dynamics of soliton formation by solving~Eq.~(\ref{1Dsgpe}) using an adaptive Runge-Kutta-Fehlberg method within the high level programming language XMDS~\cite{xmds}. 
We perform reliable simulations of the one-dimensional condensate dynamics with 8192 grid points 
on an $x$-range of $2\pi \times 512$ lattice sites. 
This corresponds to a grid spacing of 0.393, approximately three times smaller than the condensate
healing length after the solitons are formed. For our stochastic simulations we typically use 1000 trajectories and have ascertained small sampling errors (less than 5$\%$) 
in all ensemble averages.

\subsection{\label{sec:inistate}Initial state}
Present experimental techniques used to demonstrate gap solitons in a 1D optical 
lattice~\cite{Eiermann2004} allow access to the first Bloch-wave gap from the top edge
of the ground band, where the dispersion of the BEC wavepacket is negative but small. 
Therefore the strength of nonlinearity, proportional to $\gamma|\alpha|^2$, required  to
balance the dispersion and form a single soliton, is also small~\cite{Eiermann2004}. 
Hence in the experiment only very few ($\sim 350$) atoms were confined in a single
soliton.  In numerical mean-field studies of a one dimensional system~\cite{Louis2005} 
and the formation of a single gap soliton near the band edge with $\sim 100$ atoms was
shown. This result was confirmed for BECs in the presence of low-energy excitations in
the Bogoliubov approximation~\cite{Sanpera2005}.

In our analysis of a condensate with a large number of atoms at the band edge we aim  to
simulate simultaneous generation of several gap solitons in a train, thus our initial BEC 
cloud contains a few thousands of atoms if the natural scattering length of
$^{87}\mbox{Rb}$  is used~\cite{Dabrowska2006}. 
However, the truncated Wigner method requires that we add a vacuum noise contribution of half a particle per mode~\cite{Steel1998}. For a typical effective computational grid size of $4096$ points 
this gives $2048$ virtual particles. To be confident of the validity of the truncated Wigner method 
for a reasonable period of time (such that the third order derivatives in the Fokker-Planck equation 
are relatively unimportant), we wish to simulate a moderately large number of particles so that 
the coherent dynamics is not overwhelmed by the quantum noise.  

In order to enter a regime where the TWA can be successfully employed in our 1D simulations 
\emph{and} soliton trains dynamically form, in addition to having a significant number of particles, neither the effective nonlinearity nor the virtual population from vacuum noise can be too large. 
Only the condition on the nonlinearity is physical, the others are purely practical constraints arising 
from the very fine grids required for the lattice and the limitations of the TWA. 
We have found that these requirements can be reasonably met by reducing the interatomic 
interaction coupling strength to $\gamma = 10^{-4}$, corresponding  to the scattering length 
$a_s=5.4\times 10^{-11}$m, which could in principle be achieved via Feshbach resonances. 
We model the initial condensate wavefunction $\psi_0(x)$ as the Gross-Pitaevskii ground state 
for $55204$ atoms confined in a harmonic trap of frequency 
$\omega_x = 4 \pi \times 10^{-4}$Hz, giving a Thomas-Fermi radius of about $640~\mu\mbox{m}$.
The initial state of our simulations is found by solving Eq.~(\ref{1Dsgpe}) with the imaginary 
time propagation method in XMDS~\cite{xmds}.

We model the initial state of the Bose gas as a coherent state, which is realized in the Wigner 
distribution by using the initial state~\cite{Steel1998}:
\begin{align}
\label{ininoise}
\alpha(x,t=0)&=\psi_{0}(x) + \frac{1}{\sqrt{2}}\eta(x),
\end{align}
where $\psi_{0}(x)$ denotes the ground state of the condensate and $\eta(x)$ is a vacuum noise
constructed from Gaussian random variates that fulfill the conditions: 
$\langle \eta(x)\eta(x') \rangle = 0$ and $\langle \eta^{*}(x)\eta(x') \rangle = \delta(x-x')$.

For the individual simulation trajectories the initial state is expressed as
\begin{align}
\label{VN}
&\tilde{\alpha}(k,t=0)=\tilde{\psi}_{0}(k) + \frac{\theta_k}{\sqrt{2}} \; \eta(k),
\\
&
\theta_k \equiv \theta(|k|-K_{\rm max}/2),
\nonumber
\end{align}
where $\tilde{\alpha}$ denotes the spatial Fourier transform of $\alpha$, 
$K_{\rm max}$ is the maximum wave number representable on our numerical grid with 8192
points, and $\theta(k)$ is a step function. 
We note that vacuum noise is only added to half of the momentum modes 
$|k| \leq K_{\rm max}/2$. 
In our numerical simulations we ensure that significant contributions 
to the condensate evolution arises only from modes with $|k| \leq K_{\rm max}/2$, 
as modes with $|k| > K_{\rm max}/2$ are potentially affected by aliasing effects 
due to the finite extent of the grid. This procedure allows us to avoid the necessity of introducing 
a projection method for the nonlinear term into our numerical algorithm, 
e.g.~see~\cite{Bradley2005,Blakie2005}, and references within.
We have ensured in our calculations that initially empty high-momentum modes never 
have more that one tenth of the population of any other mode so that this method is justified.

\subsection{Loading the condensate onto the edge of the Brillouin zone}
\begin{figure}
\includegraphics[width=\columnwidth,keepaspectratio]{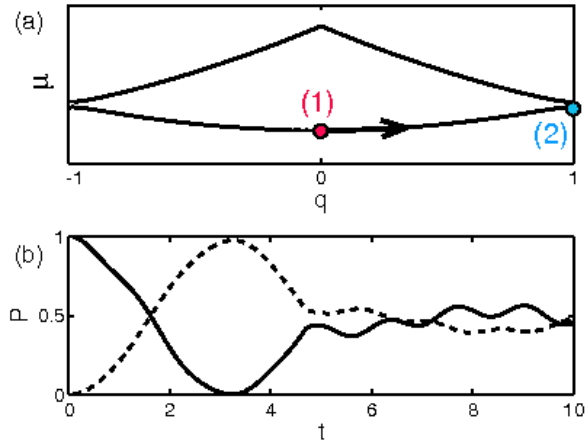}
\caption{\label{fig:rabi} 
(a) Schematics of the location (in momentum space) of the condensate wave packet relative
to the lattice band-gap structure. Points (1) and (2) correspond to the modulationally stable 
and unstable Bloch states.
(b) Rabi oscillations of the  relative population of two $k=-1$ (solid line) 
and $k=+1$ (dashed line) momentum components of a BEC nonadiabatically loaded 
into an optical lattice with the height $V_0=2$ moving with the velocity $v=1$, shown 
in the lattice rest-frame. 
The oscillations are halted by a sudden shift of the optical lattice by 
$\theta=-\pi/2$ applied at time $t_{\theta}\simeq 4.7$ ($7.48$ms), see text.
}
\end{figure}

In the mean-field approach, stationary states of a BEC are described 
by solutions of the classical Gross-Pitaevskii equation formally identical to Eq.~(\ref{1Dsgpe}), 
which take the form $$\psi_{0}(x,t) = \psi(x)\exp(-i\mu t),$$ 
where $\mu$ is the chemical potential. 
Stationary states of the non-interacting BEC loaded into a periodic potential can be expressed 
as $\psi(x)=\phi_{q}(x)\exp(i kx)$, where the wave number $k$, which in the lattice rest-frame
is equivalent to quasimomentum $q$, belongs to the first Brillouin 
zone (BZ) of the 1D lattice, and $\phi_q(x)=\phi_q(x+d)$ is a periodic (Bloch) function 
with the periodicity of the lattice. 
The spectrum $\mu(q)$ of matter-waves in an optical lattice is characterised 
by a well-known band-gap diagram as shown in Fig.~\ref{fig:rabi}(a).
Due to its periodicity in $q$ the spectrum can be reduced to the first BZ
which (in our units) extends over $-1\leq q\leq 1$. 

In order to access the regime of negative effective dispersion the condensate needs to be
prepared at the edge of the first Brillouin zone indicated by point (2) in~Fig.~\ref{fig:rabi}(a).
In our numerical simulations we implement the nonadiabatic loading technique proposed 
and experimentally demonstrated by Mellish~\emph{et al.}~\cite{Mellish2003}.
This method relies on the assumption that only the lowest energy eigenstates of the lattice
are involved in the loading process.

In the plane wave approximation the $q=0$ and $q=1$ eigenstates in the rest frame 
of the lattice are: $\phi_{q=0}(x)=(e^{-ikx}+e^{ikx})/\sqrt{2}$ and 
$\phi_{q=1}(x)=(e^{-ikx}-e^{ikx})e^{i\omega_{\small R} t}/\sqrt{2}$,
where $\omega_{\small R}$ is the Rabi frequency proportional to the energy difference 
between the states, $\phi_{q=0}(x)$ is a state from the lower edge of the band 
(point (1) in~Fig.~\ref{fig:rabi}(a)) and $\phi_{q=1}(x)$ from the upper edge
(point (2) in~Fig.~\ref{fig:rabi}(a)).
If a homogeneous condensate is suddenly loaded into an optical lattice moving with a positive 
velocity, $v>0$, it can initially be represented in the lattice rest-frame as a superposition 
of lattice eigenstates with different energies: 
$\psi_0(x)=e^{-ikx}=(\phi_{q=0}(x)+\phi_{q=1}(x))/\sqrt{2}$, and therefore will undergo Rabi 
oscillations between the $e^{ikx}$ and $e^{-ikx}$ momentum states.
After the loading, the condensate can be transfered to the $\phi_{q=1}$ state
with a sudden phase shift by  $\theta=-\pi/2$ of the lattice potential, 
$V_{\rm OL}(x)=\frac{V_0}{2}\left[1-\cos(\pi x/d -2vt+\theta)\right]$, 
applied at the time $t_{\theta}=3\pi/(2\omega_{\small R}) \simeq 4.7$ ($7.48$ms).
The applied lattice displacement is equivalent to shifting the plane wave states 
by $\theta/2$~\cite{Mellish2003}.

This reasoning can be extended to the the case of an inhomogeneous initial state of the
BEC.  As a result the Rabi oscillations are arrested as shown in~Fig.~\ref{fig:rabi}(b). 
The initial state in our simulations is suddenly released from the harmonic confinement 
into a periodic potential. After the nonadiabatic loading into an optical lattice with the height 
$V_0=2$, moving with the velocity $v=1$, the condensate undergoes Rabi oscillations, 
see Fig.~\ref{fig:rabi}(b). Following the $\theta=-\pi/2$ phase shift applied to the lattice,
the disruption of the oscillations in momentum space indicates the completion of the transfer
of the condensate to the $q=1$ momentum state, see Fig.~\ref{fig:rabi}(b).
Afterwards, the condensate is evolved for a variable time in the potential of the lattice moving 
with a velocity $v=1$ corresponding to the Brillouin zone edge.

\section{\label{sec:gapsol}Formation of gap soliton trains\protect}
\subsection{Localization signatures in the BEC density}

Within a relatively short time interval, the dynamical instability at the edge of the Brillouin zone 
starts to dominate the dynamics of the matter-wave and eventually leads to condensate localization
in the form of arrays of bright (gap) solitons. 
The development of the solitons from the bulk density  is a complex process that progresses 
over an extended period of time. Similarly to the mean-field approach~\cite{Dabrowska2006}, 
we find the development of localised soliton-like structures in the condensate density within 
the single trajectory TWA. 
As the creation of gap solitons requires relatively low nonlinearities, we take an initial wave 
packet with peak density $70$ ($5.6 \times 10^{19} \mbox{m}^{-3}$).

\begin{figure}
\includegraphics[width=\columnwidth,keepaspectratio]{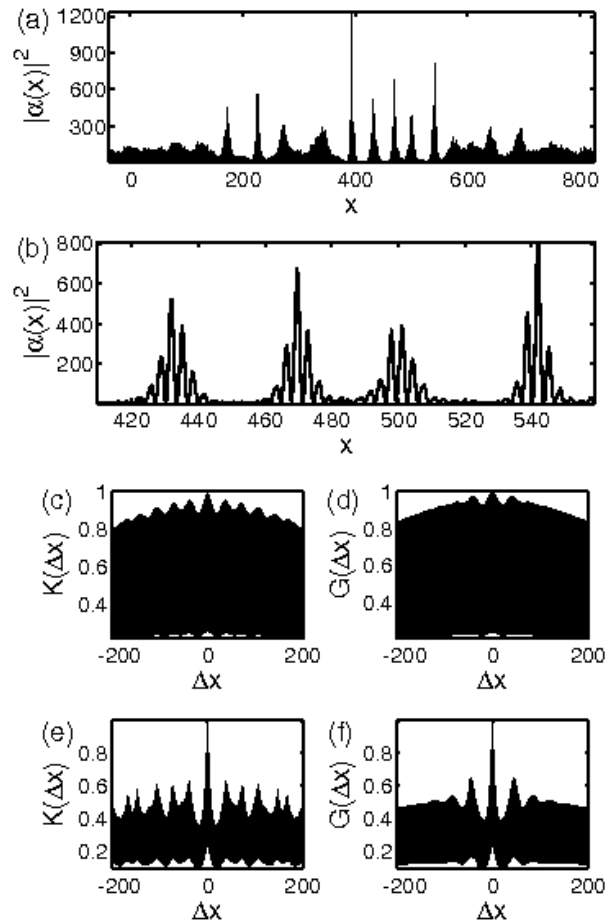}
\caption{\label{fig:D&CSing} 
(a) Density $|\alpha(x)|^{2}$ and (b) close-up of the localized structures 
in position space for a single realisation of the atom field. 
(c-f) First order density correlations in position space, see~Eqs.(\ref{Corr1},\ref{CorrM}):
(c,e) Single trajectory correlation function $K(\Delta x)$ and (d,f) integrated second order 
correlation function of the atom field averaged over 1000 trajectories $G(\Delta x)$. 
Snapshots (c,d) are taken at the time $t=266$ ($423.35$ms).
Snapshots (a,b,e,f) are taken at the time $t=406$ ($646.17$ms), 
after $t=401.3$ ($638.69$ms) of evolution at~the~BZ~edge.}
\end{figure}

For the conditions chosen in our simulations, at the time corresponding to $373.3$ 
time units ($594.13$ms) of evolution at the BZ edge, several structures with the characteristic 
beating pattern of gap solitons emerge in the central part of the cloud. 
The number of solitons developed depends on the number of atoms in the initial BEC cloud
and on the stochastic initial conditions. 
For the parameters chosen in our single-trajectory simulations, the soliton train contains 
initially seven soliton-like peaks, see Fig.~\ref{fig:D&CSing}(a).
Five solitons in a sequence are separated by a high density region from two other solitonic
structures. The high-density background encompassed by the train together with another
situated to the right of the train do give rise to further localised structures emerging at later 
evolution times. The solitons are clearly seen in the density profile $|\alpha(x)|^2$ as shown
in Fig.~\ref{fig:D&CSing}(b), and become most pronounced around the time $t=511$ ($813.28$ms).
A single soliton from the train extends over $5-7$ lattice wells. 
The centres of two neighbouring solitons in the train are separated by $\sim 12$ lattice periods. 
An average soliton contains about $2.5\times 10^3$ atoms. 

The atoms remaining in the background accumulate in the vicinity of the soliton array and exhibit 
irregular modulations of density, see Fig.~\ref{fig:D&CSing}(a). 
A small fraction of remaining atoms is scattered in the direction opposite to the movement 
of the lattice.

Once formed the solitonic structures are stable for a long time of the simulated condensate 
dynamics. In the timespan of our simulations performed for as long as $700$ time units ($1.11$s), 
up to $14$ localized soliton-like structures form. 
However at longer evolution times it becomes more difficult to clearly distinguish localised 
structures from the background density as some of the solitons are characterised by very small 
amplitudes. The solitons develop ``tails" which overlap with the background density.

The single-trajectory simulations described above are useful because they provide an
indication  of the signatures of the localization that may be seen in a ``single-shot''
realization of an experiment~\cite{Norrie2005,Norrie2006}. However, in order to 
determine populations of condensed and uncondensed atomic fractions,  the stochastic
formalism requires averaging over many trajectories. Due to the uncertainties in the
position of localized peaks introduced by the noise, the averaging of the stochastic wave
function over the ensemble does eventually lead  to the disappearance of the localized
structures in the density $\overline{\alpha^*(x) \alpha(x)}$. Since we are interested
in the enhanced thermalisation of atoms that may accompany gap soliton train formation,
we need to find another signature of the condensate localization that survives in the quantum 
ensemble average. 

\subsection{Density correlations}
A form of observable that can provide information about localization are 
density correlation functions. We define the integrated second
order correlation function as
\begin{align}
G(\Delta x) = \int_{-\infty}^{\infty} \! dx \; 
\langle \hat{\Psi}^{\dag}(x) \hat{\Psi}^{\dag}(x + \Delta x)
\hat{\Psi}(x) \hat{\Psi}(x + \Delta x)\rangle.
\label{CorrM}
\end{align}
Since we are in the classical field regime and can with some validity interpret individual
trajectories as corresponding to individual experiments, we can also define the single
trajectory correlation
\begin{align}
K(\Delta x) = \int_{-\infty}^{\infty} \! dx \; |\alpha(x+\Delta x)|^2 \; |\alpha(x)|^2,
\label{Corr1}
\end{align}
which fails to rigorously account for the vacuum occupations.  However, as these are
random and small we expect coherent features to dominate.

In Fig.~\ref{fig:D&CSing}(e) we show the shape of the correlation function for a single
trajectory,  at the time after the gap soliton train has developed. 
The second order density correlation functions $K(\Delta x)$ and $G(\Delta x)$ have been
normalised such that  their peak densities are equal to one.
They also exhibit very fast oscillations that are not resolved on the spatial scale of the figure.

The correlation function $K(\Delta x)$ in Fig.~\ref{fig:D&CSing}(e) has a characteristic central 
peak which develops only after the soliton array has formed. 
The half-width of the central peak is about the width of a single gap soliton
(roughly $\Delta x \sim 5$ lattice periods).
The development of the central peak is accompanied by the overall decay of the density 
correlations at larger $\Delta x$. 
Moreover, we observe clear regular collapses and revivals of the correlations represented by 
the sequence of maxima and minima in the envelope of $K(\Delta x)$, 
as seen in~Fig.~\ref{fig:D&CSing}(e). 
The position of the first revival ($|\Delta x| \sim 38$) coincides with the average separation distance 
between two neighbouring gap solitons in the train.

The point we stress here is that the characteristic signatures of the soliton train captured 
by the second order correlation function $K(\Delta x)$ are also present in the ensemble averaging, 
as seen in~Fig.~\ref{fig:D&CSing}(f). 
Although the exact position of the solitons in every ``single-shot" realisation of the
atom field  is uncertain due to the quantum fluctuations, matter-wave localization occurs
every time, because quantum noise always triggers dynamical instability. The localization
manifests itself in formation of soliton-like structures in the density $|\alpha(x)|^2$.
The main advantage of the density correlation function  is that it captures the average
size of these localized structures (and spacing between them) instead of the exact
location of the solitons. Hence the density correlations are less sensitive to soliton
position fluctuations between different trajectories than the averaged density. 

Similarly to the single trajectory correlation function, $G(\Delta x)$ also acquires the characteristic 
peak values when the localization occurs and the probability of finding two atoms separated 
by a certain distance in the lattice is the highest.
This happens either when $\Delta x$  is equal to zero or to the separation distance between 
centres of neighbouring solitons ($\Delta x \approx 38$).
Between the central and a neighbouring maximum of $G(\Delta x)$, a local minimum
of the correlation function at $|\Delta x| \sim 18$, which correspond to ``dips" of the matter-wave
density between two neighbouring solitons, can be seen in~Fig.~\ref{fig:D&CSing}(f).

\section{\label{sec:phase}Phase imprinting\protect}
\begin{figure}
\includegraphics[width=\columnwidth,keepaspectratio]{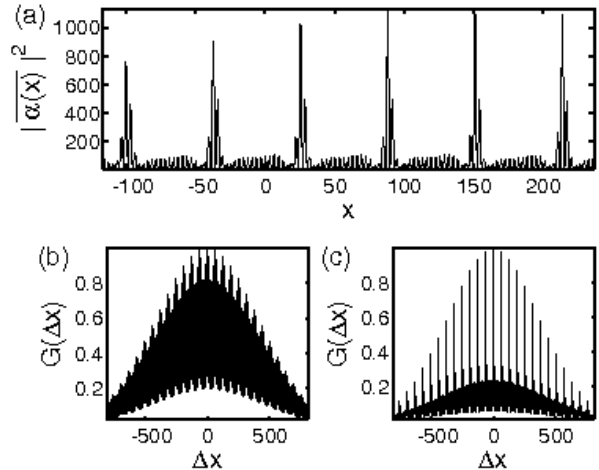}
\caption{\label{fig:MultiCorr} 
(a) Density profile $\overline{ \alpha^{*}(x)\alpha(x)}$ and (b,c) second order correlation 
functions $G(\Delta x)$, averaged over 1000 trajectories, for the case of initial phase imprinting.
Snapshots (a,c) are taken at $t=119$ ($189.39$ms), which is after $t=114.3$ ($181.91$ms) 
of evolution at the BZ edge. Snapshot of the correlation function before the gap soliton train 
develops~(b) is taken at $t=56$ ($89.13$ms).
}
\end{figure}

If the dynamical instability is seeded only by the quantum noise of the initial state, 
like in~Eq.~(\ref{ininoise}), the positions of the localised structures at each realization 
of the experiment would be random.  
If one would like to generate and manipulate the emerging solitons in a controlled fashion, 
it would be helpful to accurately forecast their position.       
This can be done by the imprinting of a periodic phase onto the initial BEC cloud {\em before} 
the condensate is loaded into a moving lattice. 
In this section, we will show that the merits of the periodic imprinting of an initial phase onto 
the condensate are two-fold. 
First, we show that the phase-imprinting leads to periodic density modulations of the condensate 
which facilitates the development of modulational (dynamical) instability and significantly reduces
the time scale of the soliton train formation. Secondly, seeding of the periodic phase (density) 
modulations leads to the development of solitons at {\em fixed locations}. 

The phase imprinting techniques, well developed for BECs~\cite{Dobrek1999,Ruostekoski2001},
were also used to place low-atom number condensates at the edge of the first Brillouin 
zone~\cite{Sanpera2005}. They result in an engineered spatially dependent 
phase factor $\exp(i\phi(x))$ of the BEC cloud. 
Here we analyse the signatures of the localization observables
if the phase of the initial BEC is regulary modulated as 
$\phi(x)=\cos(\delta \cdot x)$, where we take $\delta =0.1$ for the results we plot.
Our simulation sequence is as follows: 
First, the initial state in the harmonic trap is obtained for the same parameters as previously. 
Next, the periodic modulation of the phase is imprinted onto the BEC cloud.
Then, similarly to the previously described scheme, the real time evolution begins, 
and the condensate is nonadiabatically loaded into the moving optical lattice at the Brillouin 
zone edge. 
Finally, the condensate is left to evolve in the moving lattice for a variable time 
up to $t=700$ ($1.11$s), which is equivalent to $t=695.3$ ($1.1$s) of evolution at the BZ edge.  

As a result of the phase imprinting, strong density modulations as well as striking regularities 
within the correlation function develop at very early stages of the evolution. Periodic density 
modulation usually attributed to triggering modulational (dynamical) instabilities can be clearly 
seen already at the time $t=14$ ($22.28$ms).
Because of these, the process of the gap soliton array formation is greatly accelerated.
The solitons can be clearly visible in a single-trajectory simulation after only $t=93.3$ 
($148.49$ms) of evolution at the BZ edge.
There is also a larger number of solitons in a train (initially there are as many as $18$ localised 
structures). The number of emerging solitons and their separation 
does not depend on the period of phase modulation $\delta$ as long as the wavelength 
$\lambda=2\pi/\delta$ corresponding to the phase perturbation is larger than the characteristic 
scale for gap solitons (soliton width). 

The soliton-like structures are stable up to $t=175$ ($278.52$ms). Around $t \simeq 200$ 
their dynamics become more complicated and due to their interactions the number of solitons 
in the array changes~\cite{Dabrowska2004}.
At even longer evolution times it becomes more difficult to clearly identify localised structures 
from the background, however about $14$ soliton-like structures survive until the end 
of ``single-shot" simulations corresponding to $1.1$s.   

Most remarkably, the formation of the solitons at fixed positions manifests itself in the visibility
of the soliton train in the condensate density profile even \emph{after} ensemble averaging 
as shown in~Fig.~\ref{fig:MultiCorr}(a). The robustness of the position of the matter-wave 
localisation achieved with the aid of phase imprinting reveals itself not only in the fact that 
the solitons survive in the quantum ensemble average, but also in the much more pronounced 
periodic structure of the second order correlation function, as shown in~Fig.~\ref{fig:MultiCorr}(c) 
for the case of $1000$ trajectories.
Clear periodic revival of the second order (density) correlations can be seen.

\section{\label{sec:meanfield}Comparison with the Gross-Pitaevskii model\protect}
It has been shown~\cite{Dabrowska2006} that the nonlinear localisation of matter-waves 
due to dynamical instability at the edge of the first Brillouin zone and the development 
of gap soliton trains is a nonlinear effect which can be described by the Gross-Pitaevskii 
model, i.e. Eq.~(\ref{sgpe}) without the addition of noise to the initial state. 
Those studies were performed for a condensate \emph{adiabatically} loaded onto the edge 
of the first Brillouin zone, therefore the modeled dynamics was extended over a long period
of time. 
Nonetheless even a comparison of our stochastic simulations with the mean-field evolution
within the much faster scheme of \emph{nonadiabatic} loading shows, 
that for the same parameters the localization arises about four times earlier in the quantum 
theory. We note that if initial phase imprinting is used, it governs the time-scale
for the instability and in this scenario the train appears at similar times with and without 
the quantum noise.
We stress, that the results presented here are more accurate than simulations 
of the Gross-Pitaevskii equation, which rely on the exponential growth of numerical 
inaccuracies to show the physical effects that we have demonstrated to be in fact seeded 
by irreducible quantum noise.

\section{\label{sec:validity}Validity of simulations\protect}
Unfortunately it is difficult to estimate the simulation time for which the truncated Wigner
approximation is valid and the numerical results are quantitatively reliable.
However, for the parameters we have chosen the requirement of modeling significantly 
more particles (55000) than effective modes (4096) is fulfilled.
These numbers put our calculations into the \emph{classical field regime}, where the quantum 
noise is effectively a seed for more complicated, chaotic dynamics arising from the dynamical 
instability~\cite{Lobo2003}. These result in heating and a non-zero temperature final state,
which can not be treated by the mean-field model, see section~\ref{sec:anomalous}.
Once begun, the dynamics are dominated by nonlinear interactions between highly occupied 
modes so that even if our method gives results that are quantitatively inaccurate beyond a certain 
time, there will be a much longer time-scale for which they are qualitatively correct.
We certainly expect that the formation of soliton trains that we have demonstrated will persist 
in any experiment that can be peformed given our initial starting point.

\section{\label{sec:anomalous}Anomalous heating\protect}
\begin{figure}
\includegraphics[width=\columnwidth,keepaspectratio]{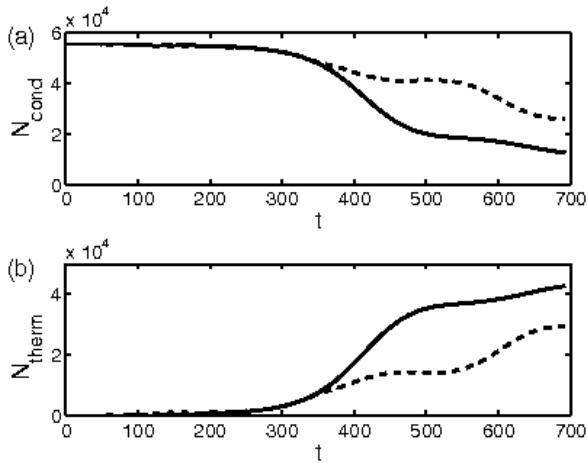}
\caption{\label{fig:C&UAtoms} 
Loss of BEC atoms~(a) and growth of the thermal fraction~(b) during $t=4.7$ 
($7.48$ms) of nonadiabatic loading followed by $t=695.3$ ($1.1$s) 
of the evolution at the BZ edge.
Shown are numbers of condensed and uncondensed atoms during the evolution 
without (solid lines) 
and with (dashed lines) the initial phase imprinting.}
\end{figure}
\begin{figure}
\includegraphics[width=\columnwidth,keepaspectratio]{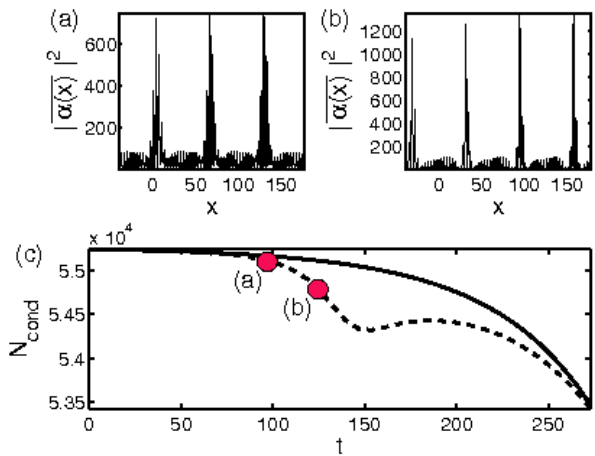}
\caption{\label{fig:C&UAtoms_Partial} 
(a,b) Close-ups of averaged density profiles $\overline{\alpha^{*}(x)\alpha(x)}$
for the scheme with addition of the phase imprinting. 
Shown are snapshots from two different time samples indicated by matching 
red circles in~(c). 
BEC profiles correspond to: (a) $t=98$ ($155.97$ms) or $93.3$ ($148.49$ms) 
of the evolution at the BZ edge, and~(b) $t=126$ ($200.54$ms) 
or $t=121.3$ ($193.05$ms) of the evolution at the BZ edge.
(c) Same as in Fig.~\ref{fig:C&UAtoms} but for shorter evolution time, 
up to $t=300$ ($477.46$ms).}
\end{figure}
As described earlier, in addition to gap soliton train formation a condensate evolving at the edge 
of the first BZ exhibits enhanced loss of atoms~\cite{Fallani2004,Inguscio2005}.
Previously~\cite{Fallani2004,Inguscio2005,Modugno2004}, the onset of the dynamical instability 
of the Bloch state at the BZ edge was conclusively linked to the growth of the thermal fraction. 
We stress, that anomalous heating can also be observed for condensates
with quasimomenta far from the BZ edge~\cite{Inguscio2005}, however, this process is triggered 
by the energetic (Landau) instabilities that are initiated by the presence of a finite thermal 
component~\cite{Inguscio2005} and facilitate decay of the condensate into the lowest energy state. 
In contrast, in our simulations the initial state is prepared as a pure coherent state BEC. 

For the parameter regime used in our simulations, the formation of gap soliton trains 
is accompanied by the enhanced thermalisation of atoms in the cloud. 
In~Fig.~\ref{fig:C&UAtoms} we present the transfer of atoms from the condensed 
to the uncondensed fraction of the Bose gas during the time evolution within our scheme.
Also shown is the thermalisation for the case of phase imprinting applied initially onto $\alpha(x)$ 
for every trajectory, see~Fig.~\ref{fig:C&UAtoms} and~Fig.~\ref{fig:C&UAtoms_Partial}.

The phase modulation results in a smaller loss of atoms from the condensate at the long time 
scales of our simulations, see~Fig.~\ref{fig:C&UAtoms}. 
However initially at the times when solitons form $t\sim 120$, it leads to an increased rate 
of atom loss, see~Fig.~\ref{fig:C&UAtoms_Partial}, which is consistent with the observation 
that the phase modulation accelerates the dynamical instability. 
It can be seen in~Fig.~\ref{fig:C&UAtoms_Partial}, that when the excitations of the condensate 
increase, in the case with initial phase modulations, the bulk density collapses and localisation commences. In~Fig.~\ref{fig:C&UAtoms_Partial}(b), showing the time sample corresponding 
to $t=126$ ($200.54$ms), the development of the soliton arrays can be clearly seen even in the averaged density profile.

\section{\label{sec:regimes}Parameter regimes for the localisation\protect}
Furthermore, we have investigated the localisation in different parameter regimes.
To this end we have monitored the condensate dynamics for higher interaction strengths: 
with coupling $\gamma$ increased to $s \gamma$, keeping all other parameters fixed. 
This reduces the peak-density at $t=0$ by a factor of $s$ 
($|\psi_{0}(t=0,x)|^{2}=\mu s^{-1} \gamma^{-1}$).
We found that for $s=10$ we do still observe the localisation of the matter-wave 
but due to the amount of noise present the pattern is much less regular than 
in the case presented in~Fig.~\ref{fig:D&CSing}.
We suspect this parameter regime to be the borderline of the applicability of the Wigner method.
We draw such a conclusion from probing the BEC dynamics for $s=100$. In this situation
we observe that any signatures of the localisation are overwhelmed by the noise.

In addition, we have investigated the dynamics with $m$ times higher nonlinearities 
but fixed densities: The coupling strength $\gamma$ and the chemical potential $\mu$
were increased by a factor of $m$ and adequately the trapping frequency of the HO
was increased by $\sqrt{m}$, setting the Thomas-Fermi radius constant.
In the case when $m=10$ our single-trajectory simulations do show a fragmentation 
of the matter-wave: We observe the development of high density spikes which we attribute 
to the anomalous heating effect in the ``single-shot'' realisation. 
Indeed in the quantum ensemble average the heating becomes more dominant.
It shows a much greater amount of thermalisation at early stages of the evolution ($t \ll 100$).
However, in the case $m=100$ thermalisation is so intensive, that no mark 
of fragmentation of the matter-wave is visible in a single-trajectory simulation. 

For $m=10$ the localisation signatures in the correlation functions ($K(\Delta x)$ and $G(\Delta x)$)
become very weak but the minima in the envelope are preserved.
On the other hand for $m=100$ all the signatures completely disappear.

\section{\label{sec:conclusions}Conclusions\protect}
Within the phase-space truncated Wigner method we have investigated the dynamics
of matter-waves in a moving lattice beyond the onset of dynamical instability.    
We have demonstrated localisation of an atomic field within the single and multitrajectory 
treatment of the TWA. 
We have shown that a train of strongly localised matter-wave gap solitons 
can be generated as a result of dynamical instability of a BEC nonadiabatically 
loaded into a moving optical lattice.
These instabilities are triggered by quantum noise.
The localization happens despite the increased thermalisation of the condensate 
at the edge of the Brillouin zone.\par

We proposed the acceleration of gap soliton array formation from moderate-size 
atomic clouds at the edge of the band by a phase imprinting technique. In addition 
we have demonstrated that the initial phase imprinting onto the BEC cloud 
leads to the localisation of matter-waves at fixed positions.
This method ensures the formation of spatially regular soliton trains at short evolution times.
Density-density correlation functions are shown to display clear signatures 
of the condensate localization.

\begin{acknowledgments}
\noindent BJDW and SW are grateful for very kind hospitality of the Quantum Atom
optics theory group at the University of Queensland.
This research was supported by the Australian Research Council (ARC)
and by a grant under the Supercomputer Time Allocation Scheme 
of the National Facility of the Australian Partnership 
for Advanced Computing~\cite{APAC}.
\end{acknowledgments}


\begin{thebibliography}{10}

\bibitem{ol_review} O. Morsch and M. Oberthaler, Rev. Mod. Phys. {\bf 78}, 179 (2006)

\bibitem{lewenstein_review}  M. Lewenstein, A. Sanpera, V. Ahufinger, B. Damski, A. S. De, U. Sen,
http://arXiv.org/abs/cond-mat/0606771 (2006)

\bibitem{Eiermann2004}
B. Eiermann, T. Anker, M. Albiez, M. Taglieber, P. Treutlein, K. P. Marzlin, and M. K. Oberthaler, Phys. Rev. Lett. {\bf 92},  230401  (2004).

\bibitem{Fallani2004}
L. Fallani, L. De Sarlo, J.E. Lye, M. Modugno, R. Saers, C. Fort, and M. Inguscio, Phys. Rev. Lett. {\bf 93},  140406  (2004).

\bibitem{Dabrowska2006}
B.~J. D\c{a}browska, E.~A. Ostrovskaya, and Y.~S. Kivshar, Phys. Rev. A {\bf 73}, 033603 (2006).

\bibitem{QN}
C.~W. Gardiner and P. Zoller, {\em Quantum Noise}, 3rd  ed. (Springer-Verlag, Berlin Heidelberg, 2004).

\bibitem{Drummond1987}
P.~D. Drummond and S.~J. Carter, J. Opt. Soc. Am. B {\bf 4},  1565  (1987).

\bibitem{Steel1998}
M.~J. Steel, M. K. Olsen, L. I. Plimak, P. D. Drummond, S. M. Tan, M. J. Collett, D. F. Walls, 
and R. Graham, {\it et~al.}, Phys. Rev. A \textbf{58},  4824  (1998).

\bibitem{Sinatra2001}
A. Sinatra, C. Lobo, and Y. Castin, Phys. Rev. Lett. \textbf{87}, 210404 (2001). 

\bibitem{Castin2002}
A. Sinatra, C. Lobo, and Y. Castin, J. Phys. B. {\bf 35}, 3599 (2002). 

\bibitem{Lobo2003} 
C. Lobo, A. Sinatra and Y. Castin, Phys. Rev. Lett. \textbf{92}, 020403 (2003).

\bibitem{Norrie2005}
A. A. Norrie, R. J. Ballagh and C. W. Gardiner, Phys. Rev. Lett. \textbf{94}, 040401 (2005). 

\bibitem{Norrie2006}
A. A. Norrie, R. J. Ballagh and C. W. Gardiner, Phys. Rev. A \textbf{73}, 043617 (2006).

\bibitem{Drummond1980}
P.~D. Drummond and C.~W. Gardiner, J. Phys. A: Math. Gen.\textbf{13}, 2353 (1980).

\bibitem{deuar2001}
P. Deuar and P. D. Drummond, Computer Physics Communications \textbf{142}, 442 (2001).

\bibitem{deuar2006}
P. Deuar and P. D. Drummond, J. Phys. A, \textbf{39}, 2723 (2006).

\bibitem{corney2004}
J. F. Corney and P. D. Drummond, Phys. Rev. Lett. \textbf{93}, 260401 (2004).

\bibitem{Polkovnikov}
A. Polkovnikov, S. Sachdev, and S.M. Girvin, Phys. Rev. A {\bf 66}, 053607 (2002)
%;A. Polkovnikov, Phys. Rev. A {\bf 68},  053604  (2003).

\bibitem{vacuum_note} The vacuum term is $1/(4h_x)$ rather than the $1/(2h_x)$ that the
reader may expect due to the fact that half of the momentum modes of the grid do not have
the initial vacuum population, see section~\ref{sec:inistate}.

\bibitem{1d_reduction} V. M. Perez-Garcia, H. Michinel, and H. Herrero, 
Phys. Rev. A {\bf 57}, 3837 (1998).

\bibitem{xmds} http://www.xmds.org

\bibitem{Louis2005}
P.~J.~Y. Louis, E.~A. Ostrovskaya, and Y.~S. Kivshar, Phys. Rev. A {\bf 71}, 023612 (2005).

\bibitem{Sanpera2005}
V. Ahufinger, and A. Sanpera Phys. Rev. Lett. {\bf 94}, 130403 (2005).

\bibitem{Bradley2005}
A. S. Bradley, P. B. Blakie and C. W. Gardiner
J. Phys. B: At. Mol. Opt. Phys. \textbf{38}, 4259 (2005).

\bibitem{Blakie2005}
P. B. Blakie and M. J. Davis,
Phys. Rev. A \textbf{72}, 063608 (2005).

\bibitem{Mellish2003}
A.~S. Mellish, G. Duffy, C. McKenzie, R. Geursen, and A.C. Wilson, Phys. Rev. A {\bf 68}, 051601 (2003).

\bibitem{Dobrek1999}
{\L}.~Dobrek, M.~Gajda, M.~Lewenstein, K.~Sengstock, G.~Birkl, W.~Ertmer, 
Phys. Rev. A {\bf 60}, 3381, (1999).

\bibitem{Ruostekoski2001}
J. Ruostekoski and J.~R. Anglin, Phys. Rev. Lett. {\bf 86}, 3934 (2001).

\bibitem{Dabrowska2004}
B.~J. D\c{a}browska, E.~A. Ostrovskaya, and Y.~S. Kivshar, Journal of Optics B: Quantum and Semiclassical Optics {\bf 6},  423 (2004).

\bibitem{Inguscio2005} L. De Sarlo, L. Fallani, J. E. Lye, M. Modugno, R. Saers, C. Fort, and M. Inguscio, Phys. Rev. A {\bf 72}, 013603 (2005)

\bibitem{Modugno2004}  M. Modugno, C, Tozzo, and F. Dalfovo, Phys. Rev. A {\bf 70}, 043625 (2004) 

\bibitem{APAC}
Details of the machine are given on the web site of the National Faciltity of
  the Australian Partnership for Advanced Computing: {\tt http://nf.apac.edu.au/}.

\end{thebibliography}
\end{document}